\documentclass[aps,twocolumn,address,abstract,showpacs,showkeys,amsmath,amssymb]{revtex4}
\usepackage {amssymb}
\usepackage {amsmath}
\usepackage{graphics,amssymb}
\usepackage{graphicx}
\usepackage{natbib}
\usepackage{revsymb}
\usepackage{bm}
\begin{document}

\title{Analytical solution of the equations describing interstitial
migration of impurity atoms}

\author{O. I. Velichko* and N. A. Sobolevskaya}

\affiliation{Department of Physics, Belarusian State University of
Informatics and Radioelectronics, 6, P.~Brovki Str., Minsk, 220013
Belarus, E-mail: oleg\_velichko@lycos.com}

\begin{abstract}
An analytical solution of the equations describing impurity
diffusion due to the migration of nonequilibrium impurity
interstitial atoms was obtained for the case of the Robin boundary
condition on the surface of a semiconductor. The solution obtained
can be useful for verification of approximate numerical solutions,
for simulation of a number of processes of interstitial diffusion,
and for modeling impurity diffusion in doped layers with the
decananometer thickness because in these layers a disequilibrium
between immobile substitutionally dissolved impurity atoms,
migrating self-interstitials, and migrating interstitial impurity
atoms can take place. To illustrate the latter cases, a model of
nitrogen diffusion in gallium arsenide was developed and
simulation of nitrogen redistribution from a doped epi-layer
during thermal annealing of $\mathrm{GaAs}$ substrate was done.
The calculated impurity concentration profile agrees well with
experimental data. The fitting to the experimental profiles
allowed us to derive the values of the parameters that describe
interstitial impurity diffusion.
\end{abstract}



\maketitle

\section{Introduction}

At present numerical methods have been widely used for simulation
of solid state diffusion of ion-implanted dopants and impurity
atoms added during epitaxial growth (see, for example,
\cite{Simulator_00,Temkin_05,Stolwijk_03}). As a rule, to simulate
the diffusion of impurity, a system of equations describing a
coupled diffusion of different mobile species and their
quasichemical reactions during annealing is solved. Due to a great
number of differential equations and the complexity of the system
as a whole, the problem of the correctness of a numerical solution
is very important. One of the best ways to verify the correctness
of the approximate numerical solutions is a comparison with the
exact analytical solution of the boundary value problem under
consideration. Such analytical solutions can be derived for the
special cases of dopant or point defect diffusion. For example, in
Ref. \cite{Minear_72} an analytical solution for the point defect
diffusion based on the Green function approach was obtained. It
was supposed in \cite{Minear_72} that nonequilibrium point defects
were continuously generated during ion implantation of impurity
atoms and diffused to the surface and into the bulk of a
semiconductor. The surface was considered to be a perfect sink for
point defects. In Ref. \cite{Velichko_88} a process of impurity
diffusion during ion implantation at elevated temperatures was
investigated analytically. It was supposed that the implantation
temperature was too low to provide a traditional diffusion by the
``dopant atom -- point defect" pairs, but was enough for the
diffusion of nonequilibrium interstitial impurity atoms to occur.
Unlike \cite{Minear_72}, in Ref. \cite{Velichko_88} a system of
equations, namely, the conservation law for substitutionally
dissolved impurity atoms and the equation of
diffusion--recombination of nonequilibrium interstitial impurity
atoms have been solved analytically by the Green function
approach. Reflecting boundary condition at the surface of a
semiconductor has been chosen to describe the interaction of
interstitial impurity atoms with the interface. Due to this
condition, a diffusion problem has become symmetric with respect
to the point $x=0$. For simplicity, the condition of zero impurity
concentration for $x \rightarrow \pm \infty$ has been used. It is
interesting to note that analytical solutions for gold diffusion
in silicon due to the Frank-Turnbull and kick-out mechanisms were
obtained in Ref. \cite{Goesele_80} and Refs.
\cite{Goesele_80,Seeger_80}, respectively. It was supposed that
there was a local equilibrium between substitutionally dissolved
impurity atoms, vacancies (or self-interstitials for the kick-out
mechanism) and interstitial impurity atoms. The case of
nonequilibrium interstitial impurity atoms was not considered in
these papers. The very interesting case of coupled diffusion of
vacancies and self-interstitials was investigated in
\cite{Hashimoto_90,Okino_95}. The equations of the diffusion of
vacancies or of self-interstitials  are similar to the equation of
diffusion of impurity interstitials. However, the solutions
obtained in \cite{Hashimoto_90,Okino_95} are difficult to use for
describing the impurity diffusion governed by nonequilibrium
interstitial impurity atoms, because a condition of local
equilibrium was used in all these papers. Besides, a generation
rate was assumed to be equal to zero in \cite{Okino_95} or equal
to a constant value in \cite{Hashimoto_90}. In
Ref.\cite{Velichko_07} analytical one-dimensional (1D) solutions
of the equations that describe impurity diffusion due to migration
of nonequilibrium impurity interstitials were obtained for the
case of impurity redistribution during ion implantation at
elevated temperatures and for diffusion from a doped epitaxial
layer. The reflecting boundary condition on the surface of a
semiconductor and the conditions of constant concentration on the
surface were used in the first and second cases, respectively.
These analytical solutions were obtained on the finite-length 1D
domain that is very convenient for comparison with numerical
solution. Moreover, simulation of hydrogen diffusion in silicon
during high-fluence low-energy deuterium implantation at a
temperature of 250 $^{\circ}$C and beryllium diffusion from a
doped epi-layer during rapid thermal annealing of InP/InGaAs
heterostructures at a temperature of 900 $^{\circ}$C was carried
out on the basis of the analytical solutions obtained. It is
interesting to note that the calculated impurity concentration
profiles agree well with the experimental data that made it
possible to derive the parameters of interstitial diffusion. This
means that the solutions obtained are useful for solving a number
of problems of solid state diffusion. The main goal of the present
work is to continue the investigations of
\cite{Velichko_88,Velichko_07} to obtain a similar analytical
solution for a more intricate case of Robin boundary condition on
the surface of a semiconductor.

\section{Original equations}

As in Refs. \cite{Velichko_88,Velichko_07}, it is supposed that
the processing temperature is too low to provide diffusion of
substitutionally dissolved impurity atoms, but is high enough for
the diffusion of impurity interstitials to occur. Nonequilibrium
interstitial impurity atoms can appear due to ion implantation or
due to the replacement of the impurity atom by self-interstitial
from the substitutional position into the interstitial one
(Watkins effect \cite{Watkins_69}), or due to dissolution of the
clusters or extended defects which incorporate impurity atoms. It
is also supposed that the concentration of impurity in the doped
regions originating from the migration of nonequilibrium impurity
interstitials is smaller or approximately equal to $n_{i}$. Here
$n_{i}$ is the intrinsic carrier concentration at the processing
temperature. If the concentration of substitutionally dissolved
dopant atoms $C$ is higher than $n_{i}$, let us confine ourselves
to the case of neutral impurity atoms in the interstitial
position. Then, the system of equations describing the evolution
of impurity concentration profiles includes
\cite{Velichko_88,Velichko_07}:

(i) the conservation law for substitutionally dissolved impurity
atoms:

\begin{equation}\label{Conservation law}
\displaystyle \frac{\partial \, C(x,t)}{ \partial \, t} =
\displaystyle \frac{C^{AI}(x,t)}{\tau^{AI}} + G^{AS}(x,t)\, ,
\end{equation}

(ii) the equation of diffusion for nonequilibrium interstitial
impurity atoms:

\begin{equation}\label{Nonequilibrium impurity interstitials}
d^{AI}\displaystyle \frac{\partial^{2} \, C^{AI}}{ \partial \,
x^{2}} - \displaystyle \frac{C^{AI}}{\tau^{AI}} + G^{AI}(x,t) = 0
\, ,
\end{equation}

or

\begin{equation}\label{Normalized equation}
- \left[ \displaystyle \frac{\partial^{2} \, C^{AI}}{ \partial \,
x^{2}} - \displaystyle \frac{C^{AI}}{l_{AI}^{2}}\right] =
\frac{\tilde{g}^{AI}(x,t)}{l_{AI}^{2}} \, ,
\end{equation}

where

\begin{equation}\label{Average migration lenght}
l_{AI}=\sqrt{d^{AI} \tau^{AI}} \, , \qquad \tilde{g}^{AI}(x,t)=
G^{AI}(x,t) \, \tau^{AI} \, .
\end{equation}

Here $C^{AI}$ is the concentration of nonequilibrium impurity
interstitials; $G^{AS}$ is the rate of adding of impurity atoms,
which immediately occupy the substitutional positions, or (with
the negative sign) the rate of the loss of substitutionally
dissolved impurity atoms due to their transfer to the interstitial
position; $d^{AI}$ and $\tau^{AI}$ are the diffusivity and average
lifetime of nonequilibrium interstitial impurity atoms,
respectively, and $G^{AI}$ is the generation rate of interstitial
impurity atoms. We use a steady-state diffusion equation for
impurity interstitials, because of the relatively large average
migration length of nonequilibrium interstitial impurity atoms
($l_{AI} \gg l_{fall}$, where $l_{fall}$ is the characteristic
length of the decrease in the concentration of substitutionally
dissolved dopant atoms) and due to the small average lifetime of
nonequilibrium impurity interstitials $\tau_{AI}$ in comparison
with the duration of thermal treatment $t_{P}$.

The system of equations (\ref{Conservation law}),
(\ref{Nonequilibrium impurity interstitials}) or
(\ref{Conservation law}), (\ref{Normalized equation}) describes
impurity diffusion due to migration of nonequilibrium interstitial
impurity atoms. To solve this system of equations, appropriate
boundary conditions are needed. Let us consider, in contrast to
\cite{Velichko_88}, the finite-length one-dimensional (1D) domain
$[0,x_{B}]$, i.e., the domain used in 1D numerical modelling. The
Robin boundary conditions on the surface of a semiconductor

\begin{equation}\label{Boundary left}
\left. \mathrm{w}^{S}_{1}d^{AI} \frac{\partial  \,
C^{AI}}{\partial x} \right |_{\displaystyle x=0} +\left.
\mathrm{w}^{S}_{2} C^{AI} \right |_{\displaystyle x=0} =
\mathrm{w}^{S}_{3} \,
\end{equation}

\noindent is added to Eq. (\ref{Normalized equation}) in
comparison with the solutions obtained in
\cite{Velichko_88,Velichko_07}. A similar condition in the bulk of
a semiconductor

\begin{equation}\label{Boundary right}
\mathrm{w}^{B}_{1}d^{AI}\left. \frac{\partial  \, C^{AI}}{\partial
x} \right |_{\displaystyle x=x_{B}} +\mathrm{w}^{B}_{2} \left.
C^{AI} \right |_{\displaystyle x=x_{B}} = \mathrm{w}^{B}_{3} \, ,
\end{equation}

\noindent as well as the initial conditions

\begin{equation}\label{Initial_conditions}
C(x,0)=C_{0}(x) \, , \qquad  C^{AI}(x,0)=C^{AI}_{eq}=const
\end{equation}

\noindent are also used for the final formulation of the boundary
value problem.

Here, $\mathrm{w}^{S}_{1}$, $\mathrm{w}^{S}_{2}$,
$\mathrm{w}^{S}_{3}$ and $\mathrm{w}^{B}_{1}$,
$\mathrm{w}^{B}_{2}$, $\mathrm{w}^{B}_{3}$ are the constant
coefficients specifying the concrete type of real boundary
conditions; $C^{AI}_{eq}$ is the equilibrium value of
concentration of interstitial impurity atoms in the bulk of a
semiconductor (it is supposed that $C^{AI}_{eq}$ is equal to zero
for many cases of interstitial impurity diffusion).

To derive an analytical solution of this boundary value problem,
the Green function approach \cite{Butkovskiy_83} can be used.

\section{Analytical method and solutions}

The suggestion about the immobility of substitutionally dissolved
impurity atoms allows one to solve Eq. (\ref{Conservation law})
independently of Eq. (\ref{Nonequilibrium impurity interstitials})
or Eq. (\ref{Normalized equation}):

\begin{equation}\label{Main integral}
C(x,t) = \frac{1}{\tau^{AI}}\int \limits_{0}^{\displaystyle
t}C^{AI}(x,t)dt +\int \limits_{0}^{\displaystyle t}G^{AS}(x,t)dt +
C_{0}(x) \, .
\end{equation}

We will use expression (\ref{Main integral}) together with the
steady-state solution of Eq. (\ref{Normalized equation}) obtained
by the Green function approach \cite{Butkovskiy_83}:

\begin{equation}\label{Steady-state_solution}
C^{AI}(x,t) = \int \limits_{0}^{\displaystyle
x_{B}}G(x,\xi)w(\xi,t)d\xi \, ,
\end{equation}

\noindent where the standardizing function $w(x,t)$
\cite{Butkovskiy_83} has the following form:

\begin{equation}\label{Standard_function}
w(\xi,t) = \frac{\tilde{g}^{AI}(\xi,t)}{l_{AI}^{2}} + w_{S}(\xi) +
w_{B}(\xi) \, .
\end{equation}

Here $G(x,\xi)$ is the Green function for Eq. (\ref{Normalized
equation}). Using the standardizing function $w(x,t)$ allows one
to reduce the previous boundary value problem to the boundary
value problem with zero boundary conditions:

\begin{equation}\label{Boundary left_zero}
\left. \mathrm{w}^{S}_{1}d^{AI} \frac{\partial  \,
C^{AI}}{\partial x} \right |_{\displaystyle x=0} +\left.
\mathrm{w}^{S}_{2} C^{AI} \right |_{\displaystyle x=0} = 0 \, ,
\end{equation}

\begin{equation}\label{Boundary right_zero}
\mathrm{w}^{B}_{1}d^{AI}\left. \frac{\partial  \, C^{AI}}{\partial
x} \right |_{\displaystyle x=x_{B}} +\mathrm{w}^{B}_{2} \left.
C^{AI} \right |_{\displaystyle x=x_{B}} = 0 \, .
\end{equation}

The Green function for Eq. (\ref{Normalized equation}) with
boundary conditions (\ref{Boundary left_zero}) and (\ref{Boundary
right_zero}) has the following form \cite{Butkovskiy_83}:

\begin{equation}\label{Green_function}
G(x,\xi)= \frac{1}{K}\left \{ \begin{array}{cc}
Q_{1}(x)Q_{2}(\xi) & \mbox{\ for } 0\leq x \leq \xi \leq x_{B} \, ,\\
\\
Q_{1}(\xi)Q_{2}(x) & \mbox{\ for } 0\leq \xi \leq x \leq x_{B} \,
,
\end{array} \right.
\end{equation}

\noindent where

\begin{equation}\label{KQ}
K= Q_{1}^{'}(x)Q_{2}(x)-Q_{1}(x)Q_{2}^{'}(x)=const \, .
\end{equation}

Here $Q_{1}$ and $Q_{2}$ are the linearly independent solutions of
the homogeneous equation

\begin{equation}\label{Homogeneous equation}
\displaystyle \frac{\mathrm{d}^{2} \, Q}{ \mathrm{d} \, x^{2}} -
\displaystyle \frac{Q}{l_{AI}^{2}} = 0 \,
\end{equation}

\noindent with the following conditions on the left boundary:

\begin{equation}\label{Boundary Q1}
Q_{1}(0)= \mathrm{w}^{S}_{1}d^{AI} \, , \qquad Q^{'}_{1}(0)=
-\mathrm{w}^{S}_{2} \, ,
\end{equation}

\noindent and on the right one:

\begin{equation}\label{Boundary Q2}
Q_{2}(x_{B})= \mathrm{w}^{B}_{1}d^{AI} \, , \qquad
Q^{'}_{2}(x_{B})= -\mathrm{w}^{B}_{2} \, .
\end{equation}

Following \cite{Butkovskiy_83}, we can write the functions
$w_{S}(x)$ and $w_{B}(x)$ as

\begin{equation}\label{wL}
   w_{S}(x)= \left \{ \begin{array}{cc}
\displaystyle
-\frac{1}{\mathrm{w}^{S}_{1}d^{AI}}\delta(-x)\mathrm{w}^{S}_{3} &
\mbox{\ if }  \, \mathrm{w}^{S}_{1} \neq 0 \, ,\\
\\
\displaystyle
\frac{1}{\mathrm{w}^{S}_{2}}\delta^{'}(-x)\mathrm{w}^{S}_{3} &
\mbox{\ if }  \, \mathrm{w}^{S}_{2} \neq 0 \, ,
\end{array} \right.
\end{equation}

\begin{equation}\label{wR}
   w_{B}(x)= \left \{ \begin{array}{cc}
\displaystyle
\frac{1}{\mathrm{w}^{B}_{1}d^{AI}}\delta(x_{B}-x)\mathrm{w}^{B}_{3}
&
\mbox{\ if }  \, \mathrm{w}^{B}_{1} \neq 0 \, ,\\
\\
\displaystyle
-\frac{1}{\mathrm{w}^{B}_{2}}\delta^{'}(x_{B}-x)\mathrm{w}^{B}_{3}
& \mbox{\ if }  \, \mathrm{w}^{B}_{2} \neq 0 \, .
\end{array} \right.
\end{equation}

Let us consider the following Robin boundary condition on the
surface of a semiconductor ($x=0$):

\begin{equation}\label{Boundary_Case left}
\left. -d^{AI} \frac{\partial  \, C^{AI}}{\partial x} \right
|_{\displaystyle x=0} + \left. \mathrm{v}^{S}_{eff} C^{AI} \right
|_{\displaystyle x=0} = 0 \, ,
\end{equation}

\noindent i.e., $\mathrm{w}^{S}_{1}=-1$, $\mathrm{w}^{S}_{2}=
\mathrm{v}^{S}_{eff}$, $\mathrm{w}^{S}_{3}=0$, and, for
simplicity, the Dirichlet boundary condition

\begin{equation}\label{Boundary_Case right}
C(x_{B},t)=C^{AI}_{B} \,
\end{equation}

\noindent in the bulk of the semiconductor, i.e.,
$\mathrm{w}^{B}_{1}=0$, $\mathrm{w}^{B}_{2}=1$,
$\mathrm{w}^{B}_{3}=C^{AI}_{B}$. Here, $\mathrm{v}^{S}_{eff}$ is
the effective escape velocity specifying the intensity of impurity
evaporation from the surface of the semiconductor.

Then, the solutions $Q_{1}$ and $Q_{2}$ have the form

\begin{equation}\label{Homogeneous solution1}
Q_{1}(x)= - d^{AI} \cosh \left (\frac{x}{l_{AI}} \right )-l_{AI}
\mathrm{v}^{S}_{eff}\sinh \left (\frac{x}{l_{AI}} \right ) \, ,
\end{equation}

\begin{equation}\label{Homogeneous solution2}
Q_{2}(x)= - l_{AI} \sinh \left (\frac{x-x_{B}}{l_{AI}} \right ) \,
\end{equation}

\noindent and

\begin{equation}\label{K}
K= - d^{AI}\cosh \left( \frac{x_{B}}{l_{AI}} \right)-l_{AI} \,
\mathrm{v}^{S}_{eff} \sinh \left (\frac{x_{B}}{l_{AI}} \right ) =
const \, ,
\end{equation}

\begin{widetext}

\begin{equation}\label{Green's_function}
   G(x,\xi)= \frac{ l_{AI}}{
   d^{AI}\cosh \left( \displaystyle \frac{x_{B}}{l_{AI}}\right)
    +l_{AI} \, \mathrm{v}^{S}_{eff}\sinh
 \left( \displaystyle \frac{x_{B}}{l_{AI}}\right)}
 \left \{ \begin{array}{cc}
 \left[d^{AI}\cosh \left( \displaystyle \frac{x}{l_{AI}}\right) +l_{AI}
  \, \mathrm{v}^{S}_{eff}\sinh \left( \displaystyle \frac{x}{l_{AI}}\right)
  \right]\sinh \left( \displaystyle \frac{x_{B} - \xi}{l_{AI}} \right )
\\
 \\
 \mbox{\ for } 0\leq x \leq \xi \leq x_{B} \, ,\\
\\
 \left[d^{AI}\cosh \left( \displaystyle \frac{\xi}{l_{AI}}\right) +l_{AI}
  \, \mathrm{v}^{S}_{eff}\sinh \left( \displaystyle \frac{\xi}{l_{AI}}\right)
  \right]\sinh \left( \displaystyle \frac{x_{B} - x}{l_{AI}} \right )
\\
\\
\mbox{\ for } 0\leq \xi \leq x \leq x_{B} \, ,
\end{array} \right.
\end{equation}
\end{widetext}

\begin{equation}\label{wL_wR}
w_{S}(x)= 0, \qquad w_{B}(x)= - \delta^{'}(x_{B}-x)C^{AI}_{B} \, .
\end{equation}

Let us consider a buried layer highly doped with impurity atoms.
Such a layer can be formed by ion implantation or due to doping
during epitaxy
\cite{Stolwijk_03,Eaglesham_94,Skarlatos_00,Mirabella_02}. If the
impurity concentration is high, a generation of nonequilibrium
interstitial impurity atoms is possible within this layer during
thermal treatment. These nonequilibrium interstitial atoms can
diffuse before they transfer to the substitutional position or are
trapped by immobile sinks. In many cases (see, for example,
impurity profiles in
\cite{Stolwijk_03,Eaglesham_94,Skarlatos_00,Mirabella_02}), the
distribution of impurity atoms in an as-grown structure can be
described by the Gaussian function

\begin{equation}\label{Generation}
C_{0}(x)=C(x,0)= C_{m}\exp \left[ -\frac{(x-R_{p})^{2}}{2\triangle
R_{p}^{ \,2}}\right] \, ,
\end{equation}

\noindent where $C_{m}$ is the maximum value of impurity
concentration; $R_{p}$ is the position of the maximum, and
$\triangle R_{p}$ is the standard deviation.

Let us suppose that the Gaussian function can also be used to
describe the spatial distribution of the generation rate of
impurity interstitials:

\begin{equation}\label{Interstitial_Generation}
G^{AI}(x,t)= g^{AI}_{m}\exp \left[
-\frac{(x-R_{p})^{2}}{2\triangle R_{p}^{ \,2}}\right] \, ,
\end{equation}

\noindent where $g^{AI}_{m}$ is the maximum rate of generation of
interstitial impurity atoms.

Taking into consideration  expressions (\ref{wL_wR}) and
(\ref{Interstitial_Generation}) yields

\begin{equation}\label{Standard_functionG}
w(\xi,t) = \frac{g^{AI}_{m}\tau^{AI}}{l_{AI}^{2}}\exp \left[
-\frac{(\xi-R_{p})^{2}}{2\triangle R_{p}^{ \,2}}\right] -
\delta^{'}(x_{B}-\xi)C^{AI}_{B} \, .
\end{equation}

Substituting the Green function (\ref{Green's_function}) and the
standardizing function (\ref{Standard_functionG}) into expression
(\ref{Steady-state_solution}) allows one to obtain a spatial
distribution of diffusing interstitial impurity atoms:

\begin{widetext}

\begin{equation}\label{Steady-state_solutionG}
\begin {array} {c}
C^{AI}(x,t) = \int \limits_{\displaystyle 0}^{\displaystyle
x_{B}}G(x,\xi)w(\xi,t)d\xi  = \displaystyle \frac{g_{m} \,
\tau^{AI}}{l_{AI}\, \displaystyle
\left[d^{AI}\cosh\left(\frac{x_{B}}{l_{AI}}
\right)+l_{AI}\mathrm{v}^{S}_{eff}\sinh\left(
\frac{x_{B}}{l_{AI}}\right)\right]}
 \\
 \\
 \times \left\{ \sinh\left( \displaystyle
\frac{x_{B}-x}{l_{AI}}\right) \displaystyle \int
\limits_{\displaystyle 0}^{\displaystyle x} \left[d^{AI}
\cosh\left( \displaystyle \frac{\xi}{l_{AI}}\right)+ \displaystyle
l_{AI}\mathrm{v}^{S}_{eff}\sinh\left( \frac{\xi}{l_{AI}}\right)
\right] \exp \left[\displaystyle
-\frac{(\xi-R_{p})^{2}}{2\triangle R_{p}^{
\,2}}\right] d\xi \right. \\
 \\
 \left. + \left[d^{AI} \cosh\left( \displaystyle
\frac{x}{l_{AI}}\right)+ \displaystyle
l_{AI}\mathrm{v}^{S}_{eff}\sinh\left( \frac{x}{l_{AI}}\right)
\right] \displaystyle \int \limits_{\displaystyle
x}^{\displaystyle x_{B}} \sinh\left( \displaystyle
\frac{x_{B}-\xi}{l_{AI}}\right) \exp \left[\displaystyle
-\frac{(\xi-R_{p})^{2}}{2\triangle R_{p}^{ \,2}}\right] d\xi
\right\} \\
 \\
 + C^{AI}_{B} \, l_{AI} \, \, \displaystyle \frac{d^{AI} \cosh\left( \displaystyle
\frac{x}{l_{AI}}\right)+ \displaystyle
l_{AI}\mathrm{v}^{S}_{eff}\sinh\left( \frac{x}{l_{AI}}\right)
}{d^{AI} \cosh\left(\displaystyle \frac{x_{B}}{l_{AI}}
\right)+l_{AI}\mathrm{v}^{S}_{eff}\sinh\left(\displaystyle
\frac{x_{B}}{l_{AI}}\right)} \int \limits_{\displaystyle
x}^{\displaystyle x_{B}} \sinh\left(\displaystyle
\frac{\xi-x_{B}}{l_{AI}}\right) \delta^{'}(x_{B}-\xi) d\xi \, .
\end {array}
\end{equation}

\end{widetext}

Calculating the integrals on the right-hand side of expression
(\ref{Steady-state_solutionG}), one can obtain an explicit
expression for the quasistationary distribution of interstitial
impurity atoms:

\begin{widetext}

\begin{equation}\label{Steady-state_solution1DG}
\begin {array} {c}
C^{AI}(x,t)= C^{AI}_{mul} \displaystyle \frac{\exp \,( u_{1})}
{d^{AI}\cosh \, u^{B}_{2}+l_{AI} \, \mathrm{v}^{S}_{eff}\sinh
 \, u^{B}_{2}} \left\{ \exp \,
(-u^{B}_{2}) \,  [\exp(u_{3})
(\mathrm{erf} \, u^{B}_{4}-\mathrm{erf} \, u_{4})\right . \\
\\
\left . +\exp(2 \,u^{B}_{2})(\mathrm{erf} u^{B}_{5}-\mathrm{erf}
\, u_{5}) \right] (d^{AI}\cosh \, u_{2}+l_{AI} \,
\mathrm{v}^{S}_{eff}\sinh
 \, u_{2})
 + \left[ \exp(u_{3})(d^{AI}+l_{AI} \, \mathrm{v}^{S}_{eff})(\mathrm{erf} \, u_{4} - \mathrm{erf} \,
u_{6}) \right. \\
\\

\left . \left .  +
 (d^{AI} - l_{AI} \, \mathrm{v}^{S}_{eff})(\mathrm{erf}
 \, u_{7}- \mathrm{erf} \, u_{5})\right] \sinh
 \, (u_{8}) \right \} +C^{AI}_{B} \displaystyle \frac{d^{AI}\cosh \, u_{2}+l_{AI} \, v^{S}_{eff}\sinh
 \, u_{2}}{d^{AI}\cosh \, u^{B}_{2}+l_{AI} \, \mathrm{v}^{S}_{eff}\sinh
 \, u^{B}_{2}} \, ,
\end {array}
\end{equation}

\end{widetext}

\noindent where

\begin{equation}\label{Cm}
C^{AI}_{mul}=\displaystyle
\frac{\sqrt{\pi}g^{AI}_{m}\tau^{AI}\Delta R_{p}}{2\sqrt{2}\,\,
l_{AI}}
 \, ,
\end{equation}

\begin{equation}\label{u1}
u_{1}= \frac{\Delta R_{p}^{2}-2l_{AI}R_{p}}{2l_{AI}^{\,2}} \, ,
\end{equation}

\begin{equation}\label{u2}
u_{2}= \frac{x}{l_{AI}}  \, ,
\end{equation}

\begin{equation}\label{uB2}
u^{B}_{2}= \frac{x_{B}}{l_{AI}} \, ,
\end{equation}

\begin{equation}\label{u3}
 u_{3}= \frac{2R_{p}}{l_{AI}} \, ,
\end{equation}

\begin{equation}\label{u4}
u_{4}= \frac{\Delta R^{\,2}_{p}+l_{AI}R_{p}-l_{AI} \, x}{\sqrt{2}
\, \Delta R_{p} \, l_{AI}} \, ,
\end{equation}

\begin{equation}\label{uB4}
u^{B}_{4}= \frac{\Delta R^{\,2}_{p}+l_{AI}R_{p}-l_{AI} \,
x_{B}}{\sqrt{2} \, \Delta R_{p} \, l_{AI}} \, ,
\end{equation}

\begin{equation}\label{u5}
u_{5}= \frac{\Delta R^{\,2}_{p}-l_{AI}R_{p}+l_{AI} \, x}{\sqrt{2}
\, \Delta R_{p} \, l_{AI}} \, ,
\end{equation}

\begin{equation}\label{uB5}
u^{B}_{5}= \frac{\Delta R^{\,2}_{p}-l_{AI}R_{p}+l_{AI} \,
x_{B}}{\sqrt{2} \, \Delta R_{p} \, l_{AI}} \, ,
\end{equation}

\begin{equation}\label{u6}
u_{6}= \frac{\Delta R^{\,2}_{p}+l_{AI}R_{p}}{\sqrt{2}\Delta
R_{p}l_{AI}} \, ,
\end{equation}

\begin{equation}\label{u7}
u_{7}= \frac{\Delta R^{\,2}_{p}-l_{AI}R_{p}}{\sqrt{2}\Delta
R_{p}l_{AI}} \, ,
\end{equation}

\begin{equation}\label{u8}
u_{8}= \frac{x-x_{B}}{l_{AI}} \, .
\end{equation}

Postulating that the loss of substitutionally dissolved impurity
atoms is equal in modulus to the rate of generation of impurity
interstitials ($G^{AS}(x,t)=-G^{AI}(x,t)$), taking into account
that the distribution of impurity interstitial atoms
$C^{AI}(x,t)=C^{AI}(x)$ for the time-independent generation rate
(\ref{Generation}), and substituting expressions
(\ref{Generation}), (\ref{Interstitial_Generation}), and
(\ref{Steady-state_solution1DG}) into (\ref{Main integral}), one
can calculate the concentration of impurity atoms in the
substitutional position

\begin{equation}\label{Main integral_calculation}
C(x,t) = \frac{t}{\tau^{AI}}C^{AI}(x) + C_{m}(1-p^{AI})\exp \left[
-\frac{(x-R_{p})^{2}}{2\triangle R_{p}^{ \,2}}\right] \, ,
\end{equation}

\noindent where $p^{AI}$ is the fraction of the impurity atoms
which transferred from the substitutional position into the
interstitial one.

Expressions (\ref{Steady-state_solution1DG}) and (\ref{Main
integral_calculation}) are the obtained solution of the boundary
value problem under consideration and can be used for verification
of approximate numerical solutions and for simulation of
interstitial diffusion.

\section{Simulation}

\begin{figure}[!ht]
\centering {
\begin{minipage}[!ht]{6.4 cm}
{\includegraphics[scale=0.58]{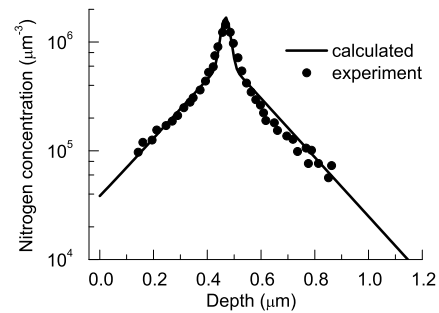}}
\end{minipage}
}

\caption{Calculated nitrogen concentration profile (solid line)
after thermal treatment of a $\mathrm{GaAs}$ substrate containing
a buried nitrogen layer with a width of several 10 nm at a
temperature of 822 $^{\circ}$C for 15 h. The experimental data
(dots) are taken from Stolwijk {\it et al.} \cite{Stolwijk_03}.
\label{fig:nitrogen}}
\end{figure}

The analytical solutions (\ref{Steady-state_solution1DG}) and
(\ref{Main integral_calculation}) can be used for modeling
different diffusion processes in semiconductor substrates. Below,
we consider the case of nitrogen diffusion in gallium arsenide
investigated in \cite{Stolwijk_03}. Nitrogen distributions of an
as-grown and subsequently annealed specimen measured by secondary
ion mass spectrometry (SIMS) are presented in
Fig.~\ref{fig:nitrogen}. In \cite{Stolwijk_03} a $\mathrm{GaAs}$
layer with about 1 $\mu$m thickness was grown on a gallium
arsenide substrate at 580 $^{\circ}$C by solid source molecular
beam epitaxy. Intermediate introduction of an $\mathrm{N}_{2}$
flow from a plasma source resulted in a buried N doping layer with
a peak concentration of about 10$^{7}$ $\mu$m$^{-3}$ and a width
of several 10 nm. This doping layer was thus sandwiched between a
buffer and a cap layer of either roughly 0.5 $\mu$m thickness.
Diffusion annealing was performed at a temperature of 822
$^{\circ}$C for 15 h in sealed quartz ampoules. It can be seen
from Fig.~\ref{fig:nitrogen} that the nitrogen concentration
profile after annealing is characterized by two extended low
concentration ``tails'' directed into the bulk of the
semiconductor and to its surface.

Let us consider a possible mechanism of the formation of such
``tails''. In recent years the mechanism of dopant diffusion in
silicon crystals due to the formation, migration, and dissociation
of the ``impurity atom -- vacancy" or ``impurity atom --
self-interstitial" pairs (the pair diffusion mechanism) has become
commonly accepted (see, for example
\cite{Simulator_00,Velichko_84,Mathiot_91,TSUPREM-4_2000}). It is
supposed within the framework of the pair diffusion mechanism that
a local thermodynamic equilibrium prevails between
substitutionally dissolved dopant atoms, intrinsic point defects,
and the pairs. However, boron diffusion in silicon is often
considered within the framework of the substitutional-interstitial
mechanism \cite{Uematsu_97,Ihaddadene_04,Martin-Bragado_04}, when
the silicon self-interstitial displaces an immobile impurity atom
from the substitutional to the interstitial position. A migrating
interstitial impurity atom in turn replaces the host atom becoming
substitutional again (the so-called ``kick-out mechanism").

It is supposed that the kick-out mechanism is also responsible for
the diffusion of gold in silicon \cite{Goesele_80,Seeger_80}; zinc
\cite{Goesele_81,Reynolds_88,Yu_91,Boesker_95,Chase_97,Chen_99,Bracht_01,Bracht_05},
nitrogen \cite{Boesker_98,Stolwijk_99,Stolwijk_03}, magnesium
\cite{Robinson_90,Robinson_91,Robinson_92}, and beryllium
\cite{Yu_91,Chen_99,Robinson_90,Hu_95} in GaAs. As follows from
the investigations presented in
\cite{Chen_99,Ketata_99,IhaddadeneLenglet_04}, beryllium diffusion
in other compound semiconductors is governed by the kick-out
mechanism too.

To describe diffusion due to the kick-out mechanism it is also
supposed that there is a local equilibrium between
substitutionally dissolved dopant atoms, self-interstitials, and
interstitial dopant atoms. In this case a mathematical description
of diffusion due to the kick-out mechanism is equal to the
description of the pair diffusion \cite{Robinson_92}. It was shown
in \cite{Velichko_88,Velichko_87,Orlowski_88} within the framework
of diffusion in silicon governed by the ``dopant atom --
self-interstitial'' pairs that the formation of the extended
``tail'' in the low concentration region of phosphorus profile
occurs if the distribution of silicon self-interstitials is
nonuniform, namely, the distribution of self-interstitials in the
neutral charge state have to be nonuniform
\cite{Velichko_88,Velichko_87}. Thus, ``tail'' formation can be
attributed to the nonuniform distribution of self-interstitials in
the neutral charge state.

In Ref. \cite{Stolwijk_03} it was supposed that nitrogen diffusion
in $\mathrm{GaAs}$ is governed by the kick-out mechanism with
interstitial $\mathrm{As}$ as a native point defect
(mathematically equivalent to the diffusion due to the ``impurity
atom -- $\mathrm{As}$ self-interstitial'' pairs), and that two
extended low concentration ``tails'' on the nitrogen concentration
profile are formed due to the nonuniform distribution of
$\mathrm{As}$ interstitial atoms. In contrast to
\cite{Stolwijk_03}, it is supposed in this paper that during
thermal annealing the generation of nitrogen interstitials occurs
in the buried layer and that migration of these nonequilibrium
interstitial atoms is responsible for the nitrogen redistribution.
In our opinion, this mechanism of diffusion is more preferable
because the atomic radius of nitrogen is smaller than arsenic
radius and hence in the doped $\mathrm{GaAs}$ the nonequilibrium
nitrogen interstitials prevail rather than the arsenic ones. In
addition, it was shown in \cite{Velichko_86} that the mass action
law and local thermodynamic equilibrium between substitutionally
dissolved dopant atoms, vacancies, and vacancy-impurity pairs are
not valid in the low concentration regions of the abrupt dopant
profile formed by low energy ion implantation. This conclusion is
also true for diffusion due to the ``dopant atom --
self-interstitial'' pairs and due to the kick-out mechanism. It is
important to note that the buried nitrogen layer investigated in
\cite{Stolwijk_03} is very narrow and disequilibrium between
substitutionally dissolved impurity atoms and diffusing species
may well occur. Therefore, we try to explain the experimental data
of \cite{Stolwijk_03} within the framework of migration of
nonequilibrium nitrogen interstitials.

In Fig.~\ref{fig:nitrogen} the results of simulation of nitrogen
diffusion in $\mathrm{GaAs}$ obtained on the basis of analytical
solution (\ref{Steady-state_solution1DG}) and (\ref{Main
integral_calculation}) are presented.

The following values of simulation parameters were used to fit the
calculated curve to the experimental nitrogen profile: i$\left .
\right)$ the parameters of the as-grown nitrogen distribution:
$C_{m}$ = 0.66$\times$10$^{7}$ $\mu$m$^{-3}$; $R_{P}$ = 0.468
$\mu$m; $\triangle R_{P}$ = 0.016 $\mu$m;  ii$\left . \right)$ the
parameters of the nitrogen interstitial diffusion: $l^{AI}$ = 0.16
$\mu$m; $\tau^{AI}$=0.1 s; $g^{AI}_m$ = 102.0
$\mu$m$^{-3}$s$^{-1}$; $p^{AI}$ = 0.84; $v^{S}_{eff}$ = 1.5
$\mu$m\,s$^{-1}$.

As can be seen from Fig.~\ref{fig:nitrogen}, the calculated curve
agrees well with the measured nitrogen concentration profile.
Thus, the experimental data \cite{Stolwijk_03} can be explained on
the basis of the migration of nonequilibrium nitrogen
interstitials. It follows from the value of the fitting parameter
$p^{AI}$ that approximately 84\% of the nitrogen atoms from the
buried layer are being transferred to the transient interstitial
positions. Migration of these nonequilibrium interstitial atoms
results in the formation of two extended ``tails" on the nitrogen
concentration profile.

\section{Conclusions}

The analytical solution of the equations that describe impurity
diffusion due to migration of nonequilibrium impurity
interstitials is obtained for the case of Robin boundary condition
on the surface of a semiconductor. Using this solution, one can
verify the correctness of the approximate numerical calculations
obtained by the codes intended for simulation of diffusion
processes. Moreover, it is possible to carry out an analytical
simulation of a number of diffusion processes which are based on
the migration of impurity interstitial atoms and can be used in
the fabrication of semiconductor devices. The solution obtained
can also be useful for simulation of impurity diffusion in the
doped layers with decananometer thickness because in these layers
disequilibrium between immobile substitutionally dissolved
impurity atoms, migrating self-interstitials, and migrating
interstitial impurity atoms can take place. As an example,
nitrogen redistribution from a buried layer during thermal
annealing of $\mathrm{GaAs}$ substrate at a temperature of 822
$^{\circ}$C for 15 h have been simulated. By comparison with the
experimental data, the values of the parameters that describe
interstitial nitrogen migration have been derived. For example, it
was found that for the process under consideration approximately
84\% of the nitrogen atoms occupied transient interstitial
positions and the average migration length of these interstitial
impurity atoms was 0.16 $\mu$m.

\end{document}